\documentstyle[12pt]{article}
\def\lsim{\mathrel{\rlap{\lower4pt\hbox{\hskip1pt$\sim$}}
    \raise1pt\hbox{$<$}}}         

  \advance\oddsidemargin  by -1in
  \advance\evensidemargin by -1in
  \advance\topmargin      by -0.0in
\makeindex

\newcommand\beq{\begin{equation}}
 \newcommand\eeq{\end{equation}}
\newcommand\beqn{\begin{eqnarray}}
 \newcommand\eeqn{\end{eqnarray}}

\def\Pom{{\bf I\!P}}

\begin{document}

\phantom{.}{\Large \hspace{10.6cm} DFTT 54/95\\
\phantom{.}\hspace{11.3cm}Torino, 1995\vspace{1.4cm}\\ }

\centerline
{\huge \bf
Color dipole systematics}
\smallskip
\centerline
{\huge \bf
of diffractive photo- and}
\smallskip
\centerline
{\huge \bf electroproduction of vector mesons}
\vspace{1.0cm}

\begin{center}
{\bf \large
J.Nemchik$^{a,b}$, N.N.~Nikolaev$^{c,d}$, E.~Predazzi$^{a}$,
B.G.~Zakharov$^{a,d}$}
\bigskip \\
{\sl
$^{a}$Dipartimento di Fisica Teorica, Universit\` a di Torino,\\
and INFN, Sezione di Torino, I-10125, Torino, Italy
\medskip\\
$^{b}$Institute of Experimental Physics, Slovak Academy of Sciences,\\
Watsonova 47, 04353 Kosice, Slovak Republik
\medskip\\
$^{c}$IKP(Theorie), KFA J{\"u}lich, 5170 J{\"u}lich, Germany
\medskip\\
$^{d}$L. D. Landau Institute for Theoretical Physics, GSP-1,
117940, \\
ul. Kosygina 2, Moscow 117334, Russia.}
\end{center}

\vspace{2.0cm}

\begin{abstract}

We present the first evaluation of color dipole cross section
from experimental data on diffractive photo- and electroproduction
of vector mesons. The dipole-size and energy dependence of
the found dipole cross section is consistent with expectations
from the BFKL dynamics.

\end{abstract}

\vspace{4.50cm}
\begin{center}
E-mail: nemchik@to.infn.it
\end{center}
\baselineskip0.785cm
\pagebreak




In the color dipole picture of high energy scattering
\cite{NZ91,NZ94,NZZ94,Mueller} the dipole cross section
is as a fundamental quantity as
the low $x$ gluon structure
function of the proton in the conventional parton model.
The diffractive electroproduction of vector mesons
\beq
\gamma^{*}p\rightarrow Vp\,,~~~~~V=\rho^{0},\,\omega^{0},\,
\phi^{0},\,J/\Psi,\,\Upsilon\,
\label{eq:1}
\eeq
offers a unique window on the dipole cross section
\cite{KZ91,NNN92,KNNZ93}. Here the
crucial point is that for the shrinkage of the transverse size
of the virtual photon with the virtuality $Q^{2}$, the
$1S$ vector meson production amplitude probes the dipole cross
section at the dipole size $r\sim r_{S}$, where
the scanning radius equals \cite{KNNZ94,NNZscan}
\beq
r_{S} = {A \over \sqrt{m_{V}^{2}+Q^{2}}}\,
\label{eq:2}
\eeq
with $A=6$.
Specifically, the amplitudes of production of the transversely
(T) and longitudinally (L) polarized $1S$ vector mesons are
of the form \cite{KNNZ94,NNZscan}
$
{\cal M}_{T} \propto r_{S}^{2}\sigma(x_{eff},r_{S}),
$
$
{\cal M}_{L} \approx {\sqrt{Q^{2}}\over m_{V}}{\cal M}_{T}
$, and one can extract the dipole cross section from the
vector meson production cross sections. For the alternative
description of vector meson production at very large
$Q^{2}$ in terms of the gluon structure function of the proton
see \cite{Ryskin,Brodsky}.

In the present communication we report the results of the
first evaluation of the color dipole cross section from
the experimental data on real photoproduction and
electroproduction of the $\rho^{0},\phi^{0}$ and $J/\Psi$
from the fixed target and HERA collider experiments. A nice
consistency with color blindness of the dipole cross section
is found. When one takes the same value of the scanning radius,
the $\rho^{0},\phi^{0}$ and $J/\Psi$ production
data yield close values of the dipole cross section.
The results of the HERA experiments
allow the first evaluation of energy dependence of the dipole
cross section at different dipole sizes and confirm the
prediction from BFKL dynamics that the smaller is the dipole
size the faster is the rise of the dipole cross section
\cite{NZZ94,NZBFKL}.

One needs the probability amplitudes to find color dipole of size $r$
in the photon and vector meson. Here
we use the formalism introduced in \cite{NZ91,KZ91} and
expounded
in \cite{NNZscan}.
The spin independence
of the dipole cross section
leads to the $s$-channel helicity conservation.
In terms of the lightcone
``radial'' wave function $\phi(r,z)$ of the $q\bar{q}$
Fock state of the vector meson
and the color dipole cross section $\sigma(\nu,r)$, the
imaginary part of the
forward production amplitudes equals
\beqn
{\rm Im}{\cal M}_{T}(x_{eff},Q^{2})=
{N_{c}C_{V}\sqrt{4\pi\alpha_{em}} \over (2\pi)^{2}}
\cdot~~~~~~~~~~~~~~~~~~~~~~~~~~~~~~~~~
\nonumber \\
\cdot \int d^{2}{\bf{r}} \sigma(x_{eff},r)
\int_{0}^{1}{dz \over z(1-z)}\left\{
m_{q}^{2}
K_{0}(\varepsilon r)
\phi(r,z)-
[z^{2}+(1-z)^{2}]\varepsilon K_{1}(\varepsilon r)\partial_{r}
\phi(r,z)\right\}\nonumber \\
 =
{1 \over (m_{V}^{2}+Q^{2})^{2}}
\int {dr^{2} \over r^{2}} {\sigma(x_{eff},r) \over r^{2}}
W_{T}(Q^{2},r^{2})
=g_{T}
\sqrt{4\pi \alpha_{em}}
C_{V}\sigma(x_{eff},r_{S}){m_{V}^{2}\over m_{V}^{2}+Q^{2}}
\, .
\label{eq:3}
\eeqn
\beqn
{\rm Im}{\cal M}_{L}(x_{eff},Q^{2})=
{N_{c}C_{V}\sqrt{4\pi\alpha_{em}} \over (2\pi)^{2}}
{2\sqrt{Q^{2}} \over m_{V}}
\cdot~~~~~~~~~~~~~~~~~~~~~~~~~~~~~~~~~
 \nonumber \\
\cdot \int d^{2}{\bf{r}} \sigma(x_{eff},r)
\int_{0}^{1}dz \left\{
[m_{q}^{2}+z(1-z)m_{V}^{2}]
K_{0}(\varepsilon r)
\phi(r,z)-
\varepsilon K_{1}(\varepsilon r)\partial_{r}
\phi(r,z)\right\}\nonumber \\
 =
{\sqrt{Q^{2}} \over m_{V}
(m_{V}^{2}+Q^{2})^{2}}
\int {dr^{2} \over r^{2}} {\sigma(x_{eff},r) \over r^{2}}
W_{L}(Q^{2},r^{2})
=
g_{L}
\sqrt{4\pi \alpha_{em}}
C_{V}\sigma(x_{eff},r_{S}){\sqrt{Q^{2}}\over m_{V}} \cdot
{m_{V}^{2}\over m_{V}^{2}+Q^{2}}\, ,
\label{eq:4}
\eeqn
where $\bf{r}$ is the transverse $q\bar{q}$ separation in the
lightcone meson, i.e., the dipole size,
$z$ is the fraction of the momentum of the meson carried by
the quark,
\beq
\varepsilon^{2} = m_{q}^{2}+z(1-z)Q^{2}\,,
\label{eq:5}
\eeq
$N_{c}=3$ is the number of colors,
$C_{V}={1\over \sqrt{2}},\,{1\over 3\sqrt{2}},\,{1\over 3},\,
{2\over 3}~~$ for the
$\rho^{0},\,\omega^{0},\,\phi^{0},\, J/\Psi$ production,
respectively, $K_{0,1}(x)$ is the modified Bessel function.
In (\ref{eq:3}),(\ref{eq:4})
 the dipole cross section enters at an effective
value of the Bjorken variable
\beq
x_{eff}= {Q^{2}+m_{V}^{2}\over 2m_{p}\nu}
\, .
\label{eq:6}
\eeq
The normalization of production amplitudes is such that
\beq
\left.{d\sigma\over dt}\right|_{t=0}={|{\cal M}|^{2}\over 16\pi}.
\label{eq:7}
\eeq

Eqs. (\ref{eq:3}),(\ref{eq:4}) give the imaginary part of production
amplitudes, one can easily include small corrections for the real
part by the substitution \cite{GribMig}
\beq
\sigma(x_{eff},r) \Longrightarrow \left(1-i
\cdot\frac{\pi}{2}\cdot\frac{\partial}
{\partial\,\log\,x_{eff}} \right)\sigma(x_{eff},r) =
\Biggl[1-i\cdot\alpha_{V}(x_{eff},r)\Biggr]\sigma(x_{eff},r)
\label{eq:8}
\eeq

Many gluon containing higher Fock states $q\bar{q}g...$ are
very important at high energy $\nu$. The crucial point is that
in leading log${1\over x}$
their effect
can be reabsorbed into the energy
dependence of $\sigma(x,r)$, which satisfies the generalized
BFKL equation (\cite{NZ94,NZZ94}, for the related approach
see also \cite{Mueller}).
The color dipole factorization formulas
(\ref{eq:3}),(\ref{eq:4})
follow from diagonalization of the scattering
matrix in the $({\bf{r}},z)$ representation.
Clearly, this  factorization holds even when the dipole
size is large, beyond the perturbative region of small
sizes. Only the wave function of the lowest
$q\bar{q}$ Fock state enters color dipole factorization
(\ref{eq:3}),(\ref{eq:4})
formulas. At large dipole size $r$ one can
identify $\phi(r,z)$, which is the probability amplitude
to find the dipole of size $r$,
with the constituent
quark wave function of the meson and to develop a viable
phenomenology of diffractive scattering which is purely
perturbative for small size mesons and allows a sensible
interpolation between soft interactions for large dipoles and
hard perturbative interactions of small dipoles. Here the
key property is the flavour independence of the pomeron
exchange dipole cross section.
The detailed description of parameterization of $\phi(r,z)$,
which incorporates the hard-QCD short distance behaviour,
is given in \cite{NNZscan}. The parameters of the wave
function were constrained to reproduce the generally
accepted radii of vector mesons and the width of leptonic
decays $V\rightarrow e^{+}e^{-}$, they are listed in Table 1.

Eqs. (\ref{eq:3}),(\ref{eq:4}) describe the pure pomeron
exchange, which dominates at large values
of the Regge parameter $\omega = 1/x_{eff}$.
At moderate values of
$\omega = 1/x_{eff}$ a substantial
part of the $\gamma p$ total cross section is due to
the non-vacuum Reggeon exchange contribution.
For instance, the Regge fit to the
$\gamma p$ total cross section can be cast in the form
\cite{DLRegge}
\beq
\sigma_{tot}(\gamma p) = \sigma_{\Pom}(\gamma p)\cdot
\left(1 + \frac{A}{\omega^{\Delta}}\right)
\label{eq:9}
\eeq
where the term $A/\omega^{\Delta}$ in the factor
$f = 1 + A/\omega^{\Delta}$ represents the non-vacuum
Reggeon exchange contribution, the Donnachie-Landshoff fit
gives $A = 2.332$ and $\Delta = 0.533$.
The similar Regge correction emerges also in real
$\rho^{0}$ photoproduction amplitude.
We do not know how large this non-vacuum contribution
to the $\rho^{0}$ production is at large $Q^{2}$,
for the crude estimation we assume that the Reggeon/pomeron
ratio scales with $\omega$, which is not inconsistent with
the known decomposition of the proton structure function
into the valence (non-vacuum Reggeon) and sea (pomeron)
contributions. Then, for the NMC kinematics we find $f = 1.25$
at $\omega \simeq 70$ relevant to $Q^{2} = 3\,GeV^{2}$ and
$f = 1.8$
at $\omega \simeq 9$ relevant to $Q^{2} = 20\,GeV^{2}$.
For the HERA energy range the non-vacuum Reggeon exchange
contribution can be neglected due to a large value
of the Regge parameter $\omega$.
In practical terms, we calculate the quantity
$d\sigma(\gamma^{*} \rightarrow V)/dt|_{t=0} =
f^{2}\cdot d\sigma_{\Pom}(\gamma^{*} \rightarrow V)/dt|_{t=0}$.
For the Zweig rule, one expects $f=1$ for the $\phi^{0},J/\Psi,
\Upsilon$ production.

In the final form of amplitudes in (\ref{eq:3}),(\ref{eq:4}) we
separated out the rapid dependence on $Q^{2}$ and/or scanning
radius $r_{S}$, and the so introduced coefficient functions
$g_{T,L}$ are smooth functions of $Q^{2}$. The possibility of
such a local
relationship between the production amplitude and the dipole
cross section at a well defined dipole size $r_{S}$ is based
on two observations:
i) the weight functions
$W_{T,L}(Q^{2},r^{2})$ have a sharp peak at
$r=A_{T,L}/\sqrt{Q^{2}+m_{V}^{2}}$ with $A_{T,L} \sim 6$
\cite{NNZscan},
ii) the ratio $\sigma(x_{eff},r)/r^{2}$ is a
smooth function of the radius \cite{NZZ94,NZBFKL}.
For these reasons, the coefficient functions $g_{T,L}$
have only a
weak sensitivity to the detailed shape of the dipole cross
section. The gross features of $\sigma(x_{eff},r)$ are well
understood and in \cite{NZHera}
good quantitative description of the small-$x$
structure function of the proton was obtained in the color
dipole BFKL dynamics.
The model dependence of $g_{T,L}$ can be
evaluated using the low-energy and high-energy forms
of $\sigma(x_{eff},r)$ described in \cite{NZHera,NNZscan},
which have a markedly different $r$ dependence.
In Fig.~1 we
present the $Q^{2}$ dependence of $g_{T,L}$ for different
production processes at $W=15$\,GeV and $W=150$\,GeV (values of
interest for the fixed target and HERA experiments respectively).
The variation of the
resulting coefficient functions $g_{T,L}$ from small to
large $W$ does not exceed $15\%$, which is
a conservative estimate of the theoretical
uncertainty of the above procedure.
The residual smooth $Q^{2}$ dependence
of $g_{T,L}$ mostly reflects the smooth and well understood
$Q^{2}$ dependence of the
scale factors $A_{T,L}$ which enter the relationship between
the position $A_{T,L}/\sqrt{Q^{2}+m_{V}^{2}}$
of the peak of $W_{T,L}(Q^{2},r^{2})$ and $r_{S}$
as given by Eq.~(\ref{eq:2}) (see also the discussion in \cite{NNZscan}).
In (\ref{eq:3}),(\ref{eq:4}), the $g_{T,L}$ are so defined as
to relate the amplitude to $\sigma(x_{eff},r_{S})$ at the
well defined scanning radius (\ref{eq:2}), reabsorbing the
effect of small departure of $A_{T,L}$ from 6 into $g_{T,L}$.

The experimentally measured forward cross production
section section equals
\beq
\frac{d\sigma(\gamma^{*} \rightarrow V)}{dt}|_{t=0} =
\frac{f^{2}}{16\pi}\cdot\Biggl[(1+\alpha_{V,T}^{2}){\cal M}_{T}^{2} +
\epsilon(1+\alpha_{V,L}^{2}){\cal M}_{L}^{2}\Biggr]
\label{eq:10}
\eeq
The difference between $\alpha_{V,L}$
and $\alpha_{V,T}$ for the longitudinal
and transverse cross sections and the overall effect of the
real part is  marginal and can safely
be neglected compared to other uncertainties.
Then,
making use of the so determined $g_{T,L}$ ,
in terms of the experimentally measured forward
production cross section we can write using Eq. (\ref{eq:3}),
(\ref{eq:4}), (\ref{eq:7}) and (\ref{eq:10})
\beqn
\sigma(x_{eff},r_{S}) = \frac{1}{f}\cdot\frac{1}{C_{V}}\cdot
{Q^{2}+m_{V}^{2} \over m_{V}^{2}}
\cdot
{2 \over \sqrt{\alpha_{em}}}\cdot
\left(
g_{T}^2 +\epsilon\,g_{L}^{2}\cdot{Q^{2}\over M_{V}^{2}}\right)^{-1/2}
\cdot     \nonumber \\
\cdot
\left(
1 +\alpha_{V}^{2}\right)^{-1/2}
\sqrt{\left.{d\sigma(\gamma^{*}\rightarrow V) \over dt}
\right|_{t=0}}
\label{eq:11}
\eeqn
Here $\epsilon$ is the longitudinal polarization of the photon
the values of which are taken from the corresponding experimental
publications.
In (\ref{eq:11}) $f$ is the above discussed factor
which accounts for the non-vacuum Reggeon contribution to
the $\rho^{0}$ production, for the $\phi^{0}$ and $J/\Psi$
production, $f \equiv 1$.
In the case the experimental data are presented in the form of
the $t$-integrated cross section, we evaluate
$
\left.{d\sigma(\gamma^{*}\rightarrow V) \over dt}
\right|_{t=0}   = B\sigma_{tot}(\gamma^{*}\rightarrow V)
$ using the diffraction slope $B$ as cited in
the same publication.
In Fig.~2  we show the results of
such analysis of the low energy \cite{Philownu}
and ZEUS \cite{ZEUSphi} data 
on real photoproduction
of the $\phi^{0}$ ,
the NMC data \cite{NMCfirho} on electroproduction of
the $\rho^{0}$ and
$\phi^{0}$, the HERA results on real and
virtual photoproduction of the $\rho^{0}$ (H1 \cite{H1rho},
ZEUS \cite{ZEUSrho94,ZEUSrho95,ZEUSrhoQ2}), fixed target data on
real photoproduction \cite{EMCPsi,E687} and electroproduction
\cite{EMCPsiQ2} of the $J/\Psi$ and HERA results on
real photoproduction
of the $J/\Psi$ (H1 \cite{H1Psi}, ZEUS \cite{ZEUSPsi}).
The error bars shown here correspond to the error bars in the
measured cross sections as cited in the experimental
publications.
The experimental data on vector meson production fall into
the two broad categories: the fixed target data taken at
typical center of mass energy $W\sim$(10-15)\,GeV and the
HERA collider data taken at $W\sim$(70-150)\,GeV. The color
dipole cross section is flavour blind, there is only kinematical
dependence on the vector meson through the definition of
$x_{eff}$. However, if different reactions are compared at
the same value of the scanning radius $r_{S}$, i.e., at
the same value of $Q^{2}+m_{V}^{2}$, then at the fixed
energy $\nu$ the corresponding values of $x_{eff}$ are equal.
Consequently, we expect that within each data group, the
procedure (\ref{eq:11}) applied to different vector mesons
will yield the same value of $\sigma(x_{eff},r_{S})$ at
the same value of $r_{S}$. This is an important consistency
check.
The experimental
data on the vector meson production give a solid evidence
for the decrease of the dipole cross section towards small
dipole size $r_{S}$. The fixed target data exhibit a decrease
of $\sigma(x_{eff},r_{S})$ by one order in the magnitude from $r_{S}
\approx 1.2$ fm in real photoproduction of the $\phi^{0}$
down to $r_{S}\approx 0.24$ fm in electroproduction
of the $\rho^{0}$ at $Q^{2}=23$\,GeV$^{2}$  and of the $J/\Psi$
at $Q^{2}=13$\,GeV$^{2}$. In the region of overlapping
values of $r_{S}$ there is a
remarkable consistency between
the dipole size dependence and also absolute
values of the
dipole cross section determined from the data on the
$\rho^{0},\phi^{0}$ and $J/\Psi$ production, in agreement with
the flavour independence of the dipole cross section.

The high energy data from HERA
exhibit a decrease of the dipole cross section
by the factor $\sim 6-7$ from $r_{S} = 1.5$\,fm
in real photoproduction of the $\rho^{0}$
down to $r_{S}= 0.26$ fm in electroproduction of the $\rho^{0}$ at
$Q^{2}=19.5$\,GeV$^{2}$.
A comparison of the fixed-target and
HERA data on real photoproduction and electroproduction
confirms the prediction \cite{NZZ94,NZBFKL,NNZscan} of
faster growth of the dipole cross section at smaller dipole
size, although the error bars are still large.

The above determination of $\sigma(x_{eff},r_{S})$ is rather
crude for the several reasons. \\
i)
First, the early EMC data on vector meson production
are well known to have been
plagued by a background from the inelastic process
$\gamma^{*}p \rightarrow VX$. A comparison of the more recent
NMC \cite{NMCfirho} and the EMC data \cite{EMCrho} on
the $\rho^{0}$ production suggests that the inelastic admixture
could have enhanced the observed cross section by as large
factor as $\sim 3$ at $Q^{2} = 17\,GeV^{2}$.
Such an uncertainty would have resulted in overestimation
of $\sigma(x_{eff},r_{S})$ by the factor $\sim 1.8$.
This may be an origin of slightly larger values of
$\sigma(x_{eff},r_{S})$ deduced from the EMC data
\cite{EMCPsiQ2} on the $J/\Psi$ electroproduction.
Also the scattering of the measured $J/\Psi$ photoproduction cross
sections is quite large, $\sim 50\%$.
Still, this factor of 2 uncertainty is much smaller
than the more than one order in magnitude variation
of $\sigma(x_{eff},r_{S})$ over the considered span in $r_{S}$.
In the recent NMC data \cite{NMCfirho} a special care has
been taken to eliminate an inelastic background and the values
of $\sigma(x_{eff},r_{S})$ from the $\rho^{0}$ and
$\phi^{0}$ production data are consistent within the experimental
error bars. \\
ii)
There are further uncertainties with the value of the
diffraction slope $B$. At large $Q^{2}$, the values of $B$
could have been underestimated due to the same inelastic
background. Even in the better quality data, there are
uncertainties with extrapolation down to $t = 0$.
Because of the curvature of the
diffraction cone, one may  somewhat
underestimate the forward cross section. The experimental
situation with the diffraction slopes is quite unsatisfactory,
in the case of the $J/\Psi$ and at large $Q^{2}$ for light
vector mesons even the $\sim 50\%$
uncertainty can not be excluded at the moment.
However, this
uncertainty in the diffraction slope corresponds to
$\lsim 25\%$ uncertainty in our evaluation of
$\sigma(x_{eff},r_{S})$, which is sufficient
for the purposes of the present exploratory
evaluation of $\sigma(x_{eff},r_{S})$. \\
iii)
There is also the above evaluated
conservative $\lsim$15\% theoretical inaccuracy of our
procedure. \\
iv) Finally, there is a residual uncertainty with the
wave function of light vector mesons. As a matter of fact,
if the dipole cross section were known, then
diffractive $\gamma^{*}p \rightarrow Vp$ must be regarded
as a local probe of the wave function of vector mesons
at $r\approx r_{S}$ \cite{KNNZ93}, which may eventually
become one of applications of vector meson production.
To this end, the consistency of $\sigma(x_{eff},r)$
determined from different reactions indicates that wave
functions of vector mesons are reasonably constrained by
the leptonic width. Here we only wish to notice, that
at high $Q^{2}$ and small scanning radius, the vector
meson production cross section is $\propto
\Gamma(V\rightarrow e^{+}e^{-})$ and even the factor
2 uncertainty in this quantity corresponds to only
$\lsim 40\%$ uncertainty in the determination of the
dipole cross section.

To summarize, within the
above stated uncertainties of our simple procedure and the
experimental error bars, there is a consistency between
the dipole cross section determined from the $\rho^{0}$
$\phi^{0}$ and
$J/\Psi$ production data.
This is the first direct determination of the dipole cross
section from the experimental data and our main conclusions
on properties of the dipole cross section are not affected
by the above cited uncertainties.

In Fig.~2 we show also the dipole cross section from the
BFKL analysis \cite{NZHera,NNZscan}, which gives a good
quantitative description of structure function of the
photon at small $x$. We conclude that the color dipole
BFKL dynamics provides a unified description of the
experimental data on diffractive production of vector
mesons and on the proton structure function.



{\bf Conclusions.}
We presented the first determination of the dipole
cross section from the experimental data on diffractive
production of vector mesons. We evaluated the color
dipole cross section  $\sigma(x_{eff},r_{S})$ at dipole size
$r_{S}$ down to $r_{S}\approx 0.2\,fm$ and confirmed the theoretically
expected rapid decrease of $\sigma(x_{eff},r_{S})$ towards small $r_{S}$.
We found a remarkable consistency
between the absolute value and the dipole size and energy
dependence of the dipole cross section extracted from the data on
different vector mesons.
This constitutes an important cross-check of the color
dipole picture of the QCD pomeron.
The found pattern of the energy dependence
of the dipole cross section is consistent
with the flavour independence and with
expectations from the BFKL dynamics.

{\bf Acknowledgements:}

This work was partly supported by the INTAS grant No. 93-239.

\pagebreak

\pagebreak

\begin{table}[h]
\begin{center}
\begin{tabular}{||l||c|c|c|c|} \hline\hline
 parameter   & $\rho^{0}$ & $\phi^{0}$ & $J/\Psi$ & $\Upsilon$ \\ \hline
$R^2 [fm^2]$ &    1.370   &    0.690   &  0.135   & 0.015      \\ \hline
$C$          &    0.360   &    0.530   &  1.130   & 1.990      \\ \hline
$m_q$ [GeV]  &    0.150   &    0.300   &  1.300   & 5.000      \\ \hline\hline
\end{tabular}
\end{center}
\caption[.]{ \sl Values of the parameters $R^2$, $C$ and $m_q$
of the wave function of vector mesons \cite{NNZscan}.}
\end{table}

\pagebreak
{\bf Figure captions:}
\begin{itemize}

\item[Fig.~1]
The $Q^{2}$ dependence of the coefficient functions
$g_{T,L}$ at $W=15$ GeV (dashed curve) and $W=150$ GeV
(solid curve).

\item[Fig.~2]
The dipole size dependence of the dipole cross section
extracted from the experimental data on photoproduction
and electroproduction of vector mesons:
the NMC data on $\phi^0$ and $\rho^0$ production \cite{NMCfirho},
the EMC data on $J/\Psi$ production \cite{EMCPsi,EMCPsiQ2},
the E687 data on $J/\Psi$ production \cite{E687},
the FNAL data on $\rho^0$ production \cite{Philownu},
the ZEUS data on $\rho^0$ production \cite{ZEUSrho94,ZEUSrho95,ZEUSrhoQ2}, 
the ZEUS data on $\phi^{0}$ production \cite{ZEUSphi},
the H1 data on $\rho^0$ production \cite{H1rho} and
the average of the H1 and ZEUS data on $J/\Psi$ 
production \cite{H1Psi,ZEUSPsi}.
The dashed and solid curve show the dipole cross section
of the model \cite{NZHera,NNZscan} evaluated for the c.m.s. energy
$W=15$ and $W=70$ GeV respectively.
The data points at HERA energies and the corresponding solid curve
are multiplied by the factor 1.5.
\end{itemize}

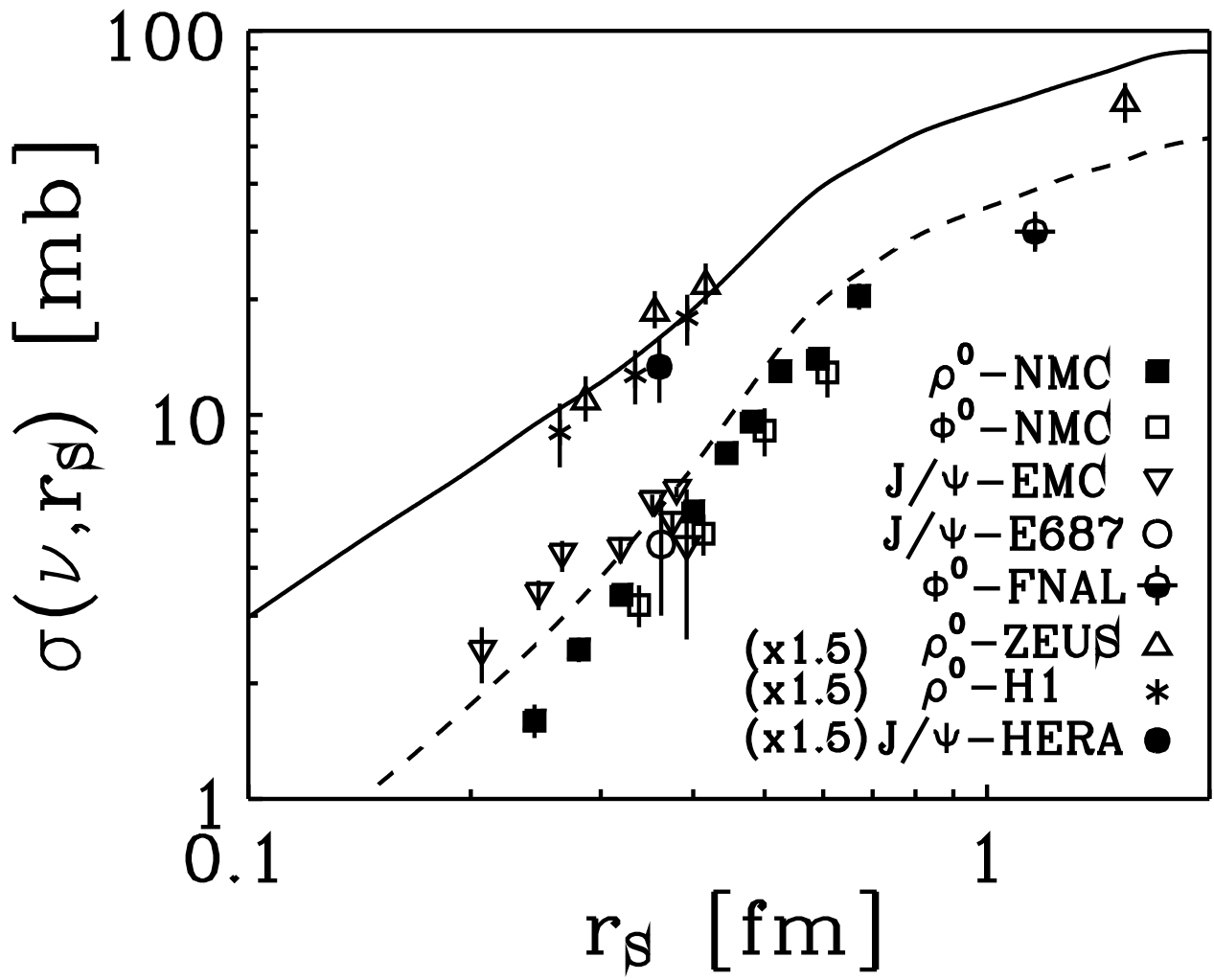
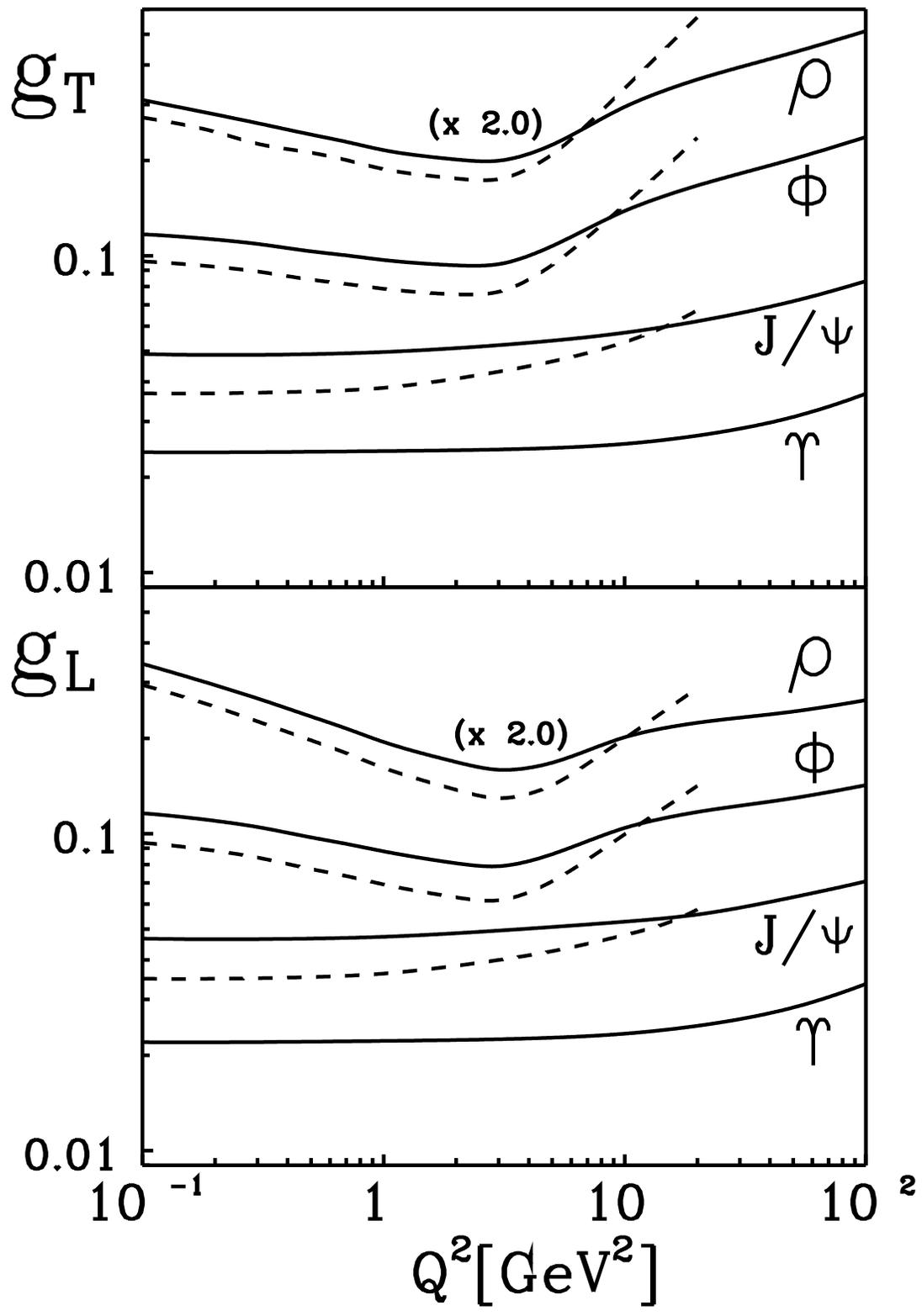
\end{document}